# Interference Declination for Dynamic Channel Borrowing Scheme in Wireless Networks

Shakil Ahmed, Mohammad Arif Hossain, and Mostafa Zaman Chowdhury
Department of Electrical and Electronic Engineering
Khulna University of Engineering & Technology, Khulna-9203, Bangladesh
E-mail: shakileee076@gmail.com, dihan.kuet@gmail.com, mzceee@yahoo.com

***Abstract*— In modern days, users in the wireless networks are increasing drastically. It has become the major concern for researchers to manage the maximum users with limited radio resource. Interference is one of the biggest hindrances to reach the goal. In this paper, being deep apprehension of the issue, an efficient dynamic channel borrowing scheme is proposed that ensures better Quality of Service (QoS) with interference declination. We propose that if channels are borrowed from adjacent cells, cell bifurcation will be introduced that ensures interference declination when the borrowed channels have same frequency band. We also propose a scheme that inactivates the unoccupied interfering channels of adjacent cells, instead of cell bifurcation for interference declination. The simulation outcomes show acceptable performances in terms of SINR level, system capacity, and outage probability compared to conventional scheme without interference declination that may attract the considerable interest for the users.**

## I. Introduction

Wireless networks are searching for an ease way to keep pace with the fastest growing users with maximum Quality of Service (QoS). As the radio resource is limited and which may cause interference problem in case of same frequency allocation, maximum utilization of radio resources with interference declination is the only way to quench the thirst of the users. In a network, instead of huge traffic, there may be unoccupied channels for particular cells. If a scheme can be developed in such a way where maximum utilization with interference declination for the unoccupied channels is ensured, these channels can be appealed for the users. For a cell where the traffic intensity is higher i.e. the number of users is greater than the total channels, the cell can take the advantage by borrowing channels from adjacent cells [1]-[3]. When channels are borrowed, there may be huge interference from adjacent cells because of reused frequencies. In this paper, we propose a scheme where the bifurcation of cells will be done so that the same frequency band of adjacent cells can be resided for the inner part users. We propose a unique way that ensures the inactivation of unused channels of same frequencies for the adjacent cells.

In the previous time, fixed channel assignment (FCA) [4] and hybrid channel assignment (HCA) [5] have been proposed to utilize the bandwidth without considering interference declination. Dynamic channel allocation with interference mitigation architecture [6] and interference declination approach for OFDMA networks [7], [8] describe some interference management techniques. Our proposed model shows better performances compared to the above models.

In the proposed scheme, the cell bifurcation may be done for adjacent cells (from where channels are not borrowed) having same frequency spectrum and inactivating the interfering channels of the adjacent cells, if there are any unoccupied channels. It results in increased bandwidth utilization with interference declination, higher capacity, and lower outage probability for a cellular network where there is excessive traffic.

The rest of the paper is systematized as followed: Section II shows the proposed interference declination scheme with proper illustration. The capacity and the outage probability analysis for the proposed scheme are shown in Section III. Section IV demonstrates the simulation results for the proposed scheme. Finally, conclusions are drawn in Section V.

## II. Proposed Interference Declination Scheme

Modern and forthcoming wireless network should be promising to assist QoS to its users. Interference declination is one of the ways to serve better QoS. It can make a wireless network more worthy to users.

### *A. System model*

In our analysis, we consider a cluster of seven cells. So, there are three types of reused frequency band. Let, the frequency bands are named as *A*, *B*, *C*. Fig. 1 reflects the frequency allocation of cells before channel borrowing of the system. Bandwidth spectrum of corresponding Fig. 1 has been depicted in Fig. 2. According to our assumptions, cell 1 is the reference cell where the traffic intensity is higher than other six cells of the cluster. As a result, there creates an inevitability to borrow channels from adjacent cells. As the frequency bands are reused, there exists huge probability to receive interference from the adjacent and neighboring cells. When the reference cell requires channels, it can borrow required number of channels from the adjacent cells in such a way that no interference may occur. The channel borrowing process has been shown in Fig. 3 which refers that cell 1 borrowed $B'$ and $C'$ channels from cell 2 and cell 3, respectively. Consequently, interference declination becomes obvious as cell 1 may get interference from cell 4 and cell 6 for $B'$ channels or from cell 5 and cell 7 for $C'$ channels, respectively. Fig. 4 depicts the bandwidth spectrum of corresponding state after borrowing channels. However, when interference mitigation is required, two different techniques have been proposed to reduce interferences.

### B. *Keeping interfering channels blocked*

The novelty of the proposed scheme points towards the process of making the interfering channels of the adjacent cells inactive when the reference cell borrows channels from other adjacent cells. Fig. 5(a) illustrates the appliance for interference mitigation. When the reference cell (cell 1) borrows channels ($B'+C'$) from adjacent cells (cell 2 and cell 3), the interfering channels has been managed in such a way that reduces interference.

When $B'$ and $C'$ channels are borrowed by the reference cell from cell 2 and cell 3, respectively, it may possibly receive interference from the same channels of the other adjacent cells. According to the manner, if the interfering channels are unoccupied, they can be blocked. As a result, interference can be reduced. Therefore, B' channels of cell 4 and cell 6 have to keep blocked as well as C' channels of cell 5 and cell 7, respectively. Consequently, interference to the users of the cluster can be reduced to a great extent.

### C. *Bifurcation of adjacent cells*

Another approach for interference declination has been proposed in which the adjacent cells have to be bifurcated that diminishes the sophistication of the system with creditable performance. If the interfering channels are occupied in the adjacent cells, then bifurcation technique has been applied. The bifurcation process has been demonstrated in Fig. 5(b). In this process, if $B'$ and $C'$ channels are borrowed by the reference cell from cell 2 and cell 3, respectively, then the reference cell and other cells are bifurcated in such a way that interference can be mitigated.

According to the system model, $B'$ and $C'$ channels are allocated to the inner part users of the reference cell. Next step comprises with $B''$ *and* $B'''$ channels that have been provided according to requirement of the inner users of cell 4 and of cell 6, respectively, where $0 \leq B'' \leq B'$ and $0 \leq B''' \leq B'$. Similarly, $C''$ and $C'''$ channels are provided according to requirement of the inner users of cell 5 and cell 7, respectively, where $0 \leq C'' \leq C'$ and $0 \leq C''' \leq C'$. Here, $B''$ and $B'''$ indicate the portion of occupied interfering channels. When inner users of cell 4 and cell 6 need more channels, then $B''$ and $B'''$ can be extended up to $B'$, respectively.

Similar process can be applicable for cell 5 and cell 7. When inner par users of cell 5 and cell 7 need more channels to keep the newly arrived traffic active, $C''$ and $C'''$ channels can be extended up to $C'$ channels, respectively. If the traffic cannot fill up $C'$, then some portion of $C'$ remains unused. It the number of traffic exceeds $C'$, then these additional calls will be blocked. The mitigation process has been demonstrated in Fig. 6.

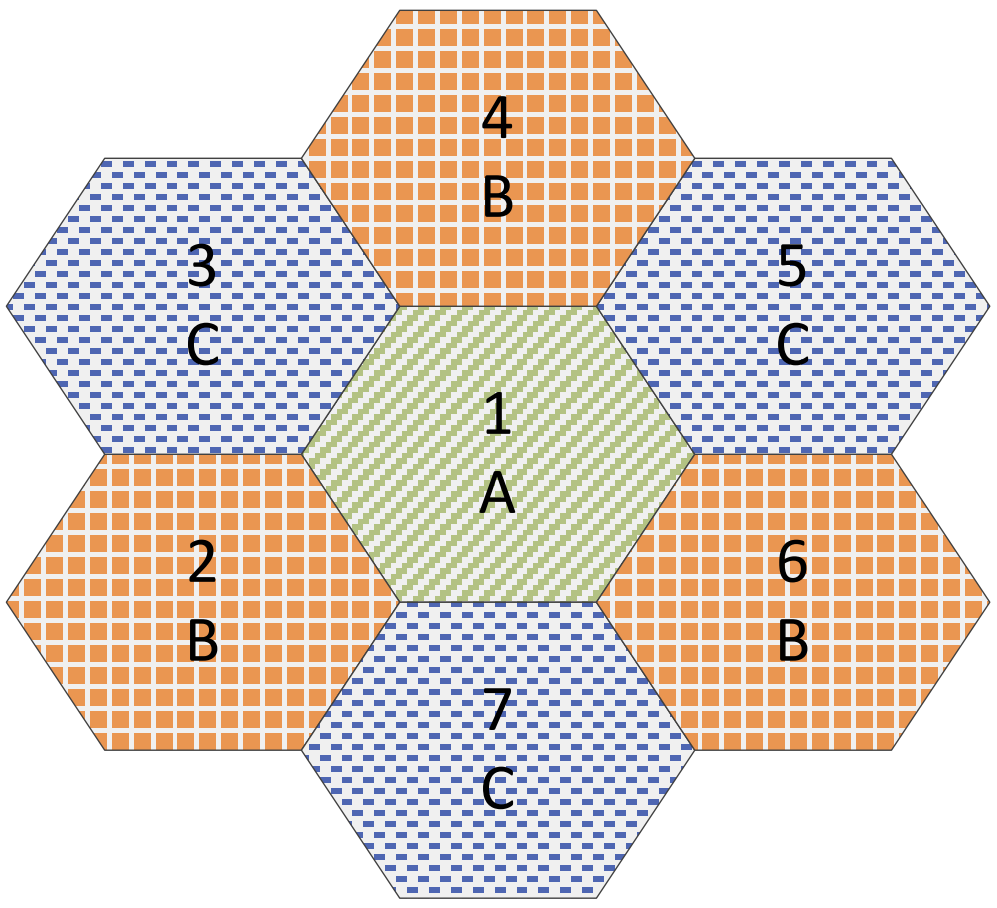

**Fig. 1.** Frequency band allocation before channel borrowing.

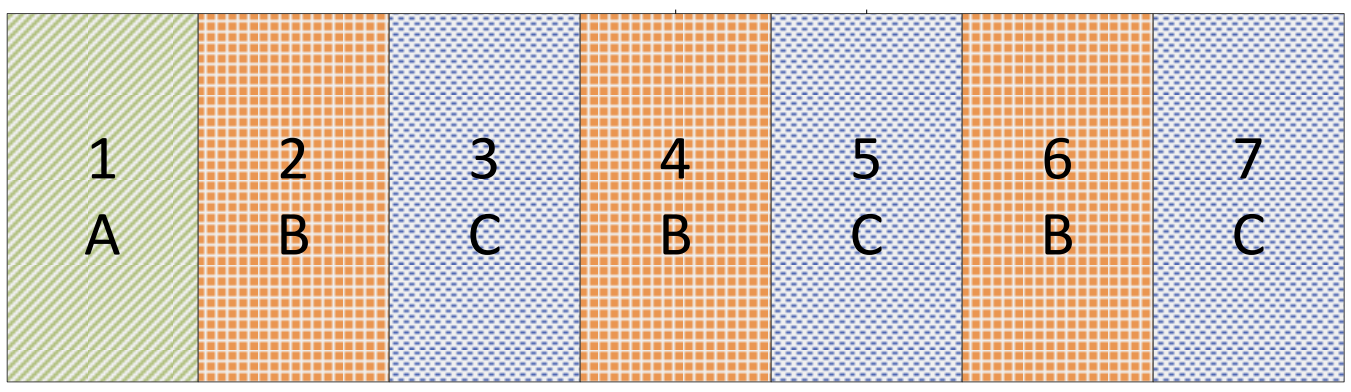

**Fig. 2**. Frequency spectrum before channel borrowing.

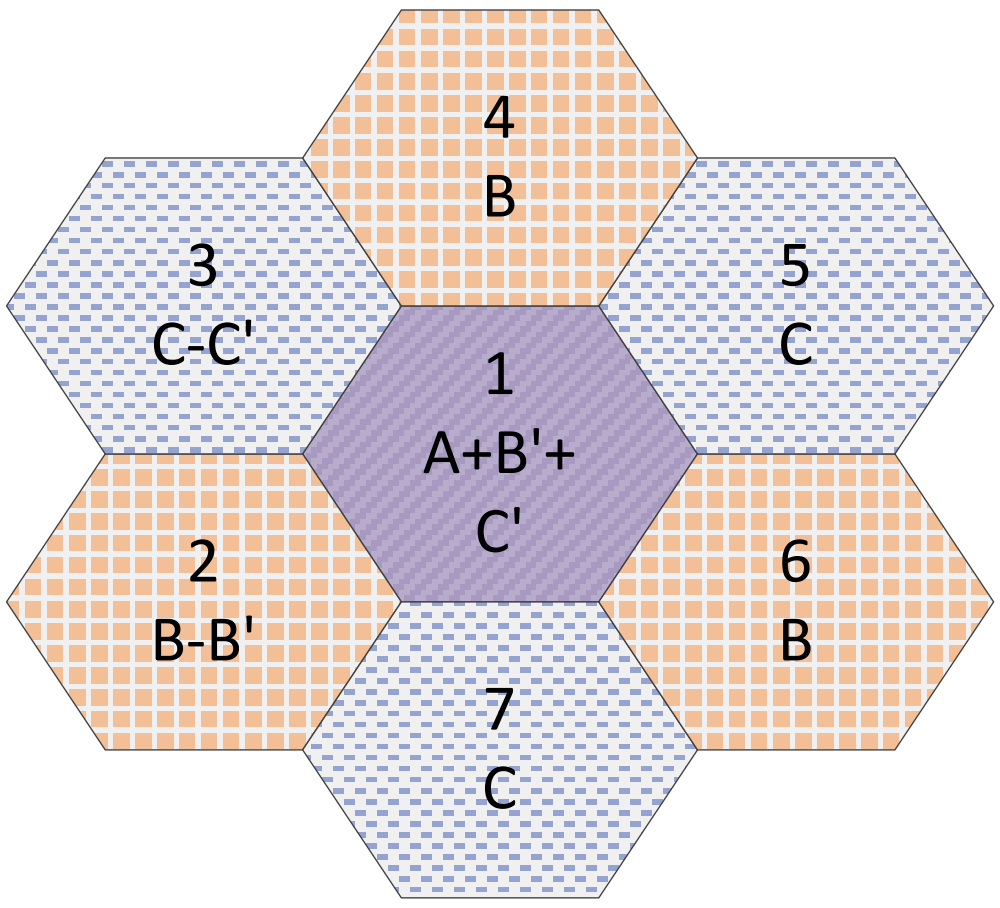

**Fig. 3**. Frequency allocation after channel borrowing process without interference declination.

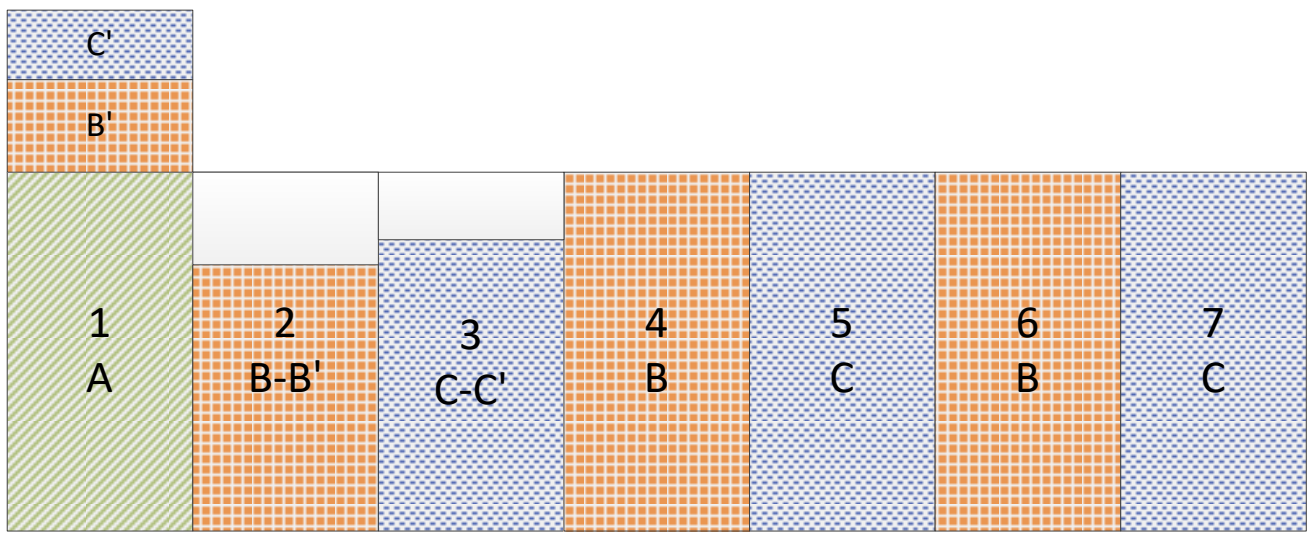

**Fig. 4**. Frequency spectrum after channel borrowing.

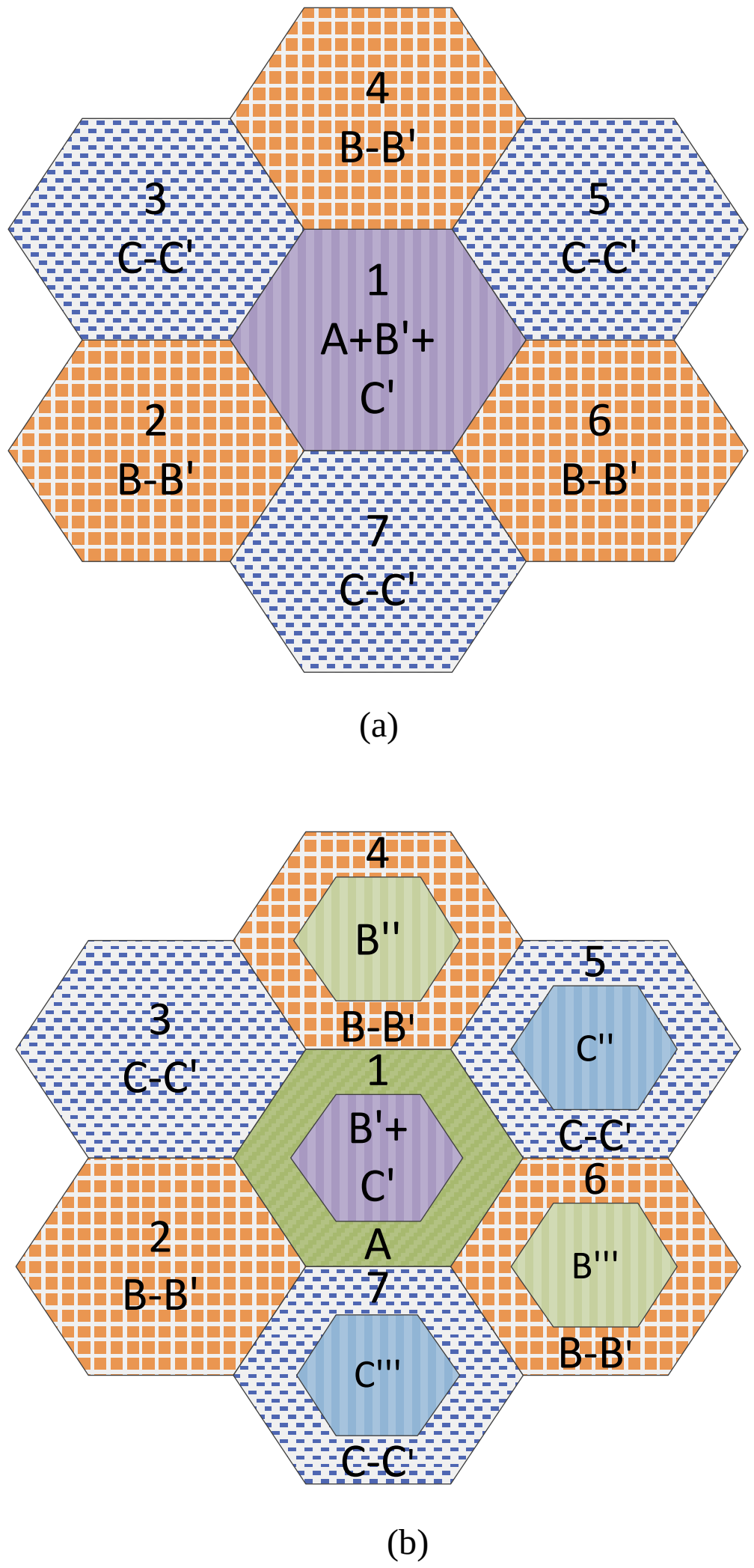

**Fig. 5.** Frequency allocation after channel borrowing process with interference declination (a) Blockage of the interfering channels (b) Bifurcation process of adjacent cells.

## III. Capacity and Outage Probability Analysis for the Proposed Scheme

For interference declination, there are many mechanisms available at present in the cellular networks. Optimization of interference and noise is one of the biggest challenges because these severely increase the outage probability as well as decrease the capacity of the wireless link.

We use Okumura-Hata model for macro-cellular path calculation [9] as

$$L = 69.55 + 26.16 \log f_c - 13.82 \log h_b - a(h_m) + (44.9 - 6.55 \log h_b) \log d \quad (1)$$

$$a(h_m) = 1.1(\log fc - 0.7)h_m - (1.56 \log fc - 0.8) \quad (2)$$

where $f_c$ is the centre frequency of the cell, $h_b$ is the height of base station (BS) of the cell, $h_m$ is the height of user (mobile antenna), $d$ is defined as the distance between the BS and the user.

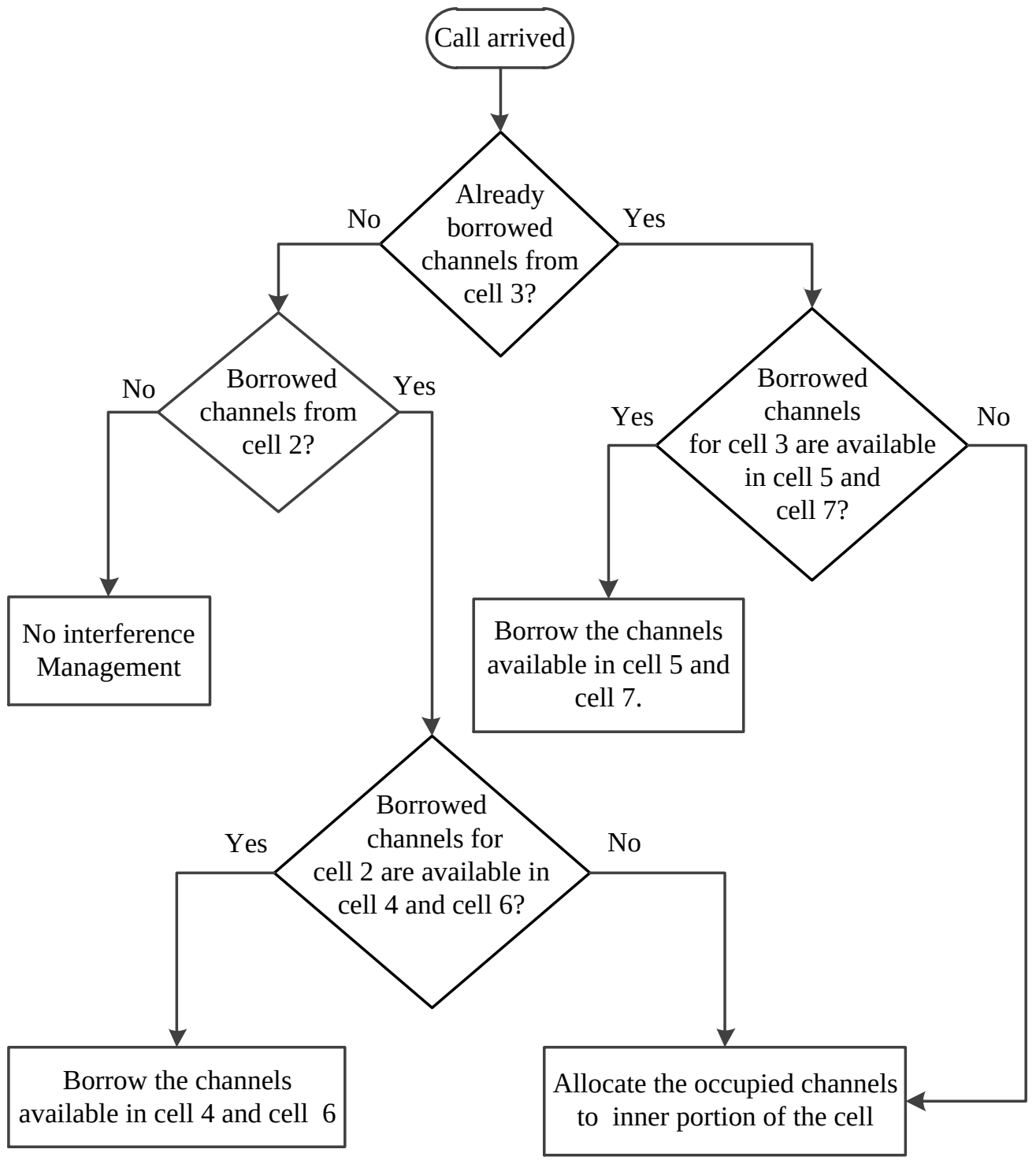

**Fig. 6.** Interference declination for adjacent cell bifurcation.

In case of capacity, we can write Shannon capacity formula

$$C = \log_2(1 + SINR) \ \ [\text{bps/Hz}] \quad (3)$$

The capacity of a wireless network is related to signal to interference plus noise ratio (SINR) level. The received SINR level for the users can be expressed as

$$SINR = \frac{S_0}{\sum_{i=1}^{M} I_i + \sum_{k=1}^{P} I_k + N_0} \quad (4)$$

where $S_0$ is the signal power of the BS, $N_0$ is total power of the received noise, $M$ and $P$ are the number of interfering cells in 1st tier and 2nd tier, respectively according to the proposed model.

The outage probability [10], [11] can be calculated as

$$P_{out} = \Pr\ (SINR < \gamma) \quad (5)$$

where $\gamma$ is the threshold value of SINR under which there is no acceptable response.

For the proposed model, considering the interfering adjacent cells, the outage probability can be expressed as

$$P_{out} = 1 - \prod_{i=1}^{M+P} \exp(-\frac{\gamma}{S} I_i) \quad (6)$$

The bifurcation of cell reduces the outage probability ($P_{out}$), improves the SINR level and the capacity of the system.

## IV. SIMULATION RESULTS

In this section, we have appraised SINR level, system capacity, and the outage probability for the proposed dynamic channel borrowing scheme. We summarize the assumptions that are required for the simulation of the proposed scheme in Table. I. We have considered 1st and 2nd tier of the reference cell for our simulation.

Fig. 7 shows the SINR level for the proposed scheme and conventional scheme i.e. scheme without interference declination after dynamic channel borrowing process. The result demonstrates such performance that is in favor of the proposed scheme. It implies that the SINR level decreases with the increment of the distance between the base station (BS) and the user of the reference cell for both schemes. The interference management process increases the SINR level which is very noteworthy for any distance compared to the conventional scheme. As bifurcation is done and the interfering channels are provided to the inner part users, they get strong signal from the BS of the reference cell and small interfering signal from the BS of the adjacent cells.

Fig. 8 shows the comparison of the capacity between the proposed scheme and the conventional scheme with no interference declination after dynamic channel borrowing process. Careful study shows that the proposed scheme affords significant performance compared to conventional scheme. The figure clearly shows that the proposed scheme increases the capacity of the network. It also shows that the capacity decreases with the increment of distance between the BS and the user of the reference cell.

Fig. 9 illustrates the fact that the outage probability of the proposed scheme is meaningfully smaller than before interference declination. With the increment of distance between BS and users, the outage probability of the proposed scheme decreases more as compared to conventional scheme. In short, the proposed scheme points toward the reduction in SINR level and outage probability along with the aggrandizement of the capacity of cellular network.

TABLE I: Summary of the parameter values used in analysis

| Consideration | Value |
|---|---|
| Number of cell in each cluster | 7 |
| Number of reused frequency | 3 |
| Threshold value of SINR( $\gamma$ ) | 9 dB |
| Centre frequency | 1800 MHz |
| Transmit signal power by the BS | 1.50 kw |
| Height of the BS | 100 m |
| Height of mobile antenna | 5 m |
| Cell radius | 1 km |

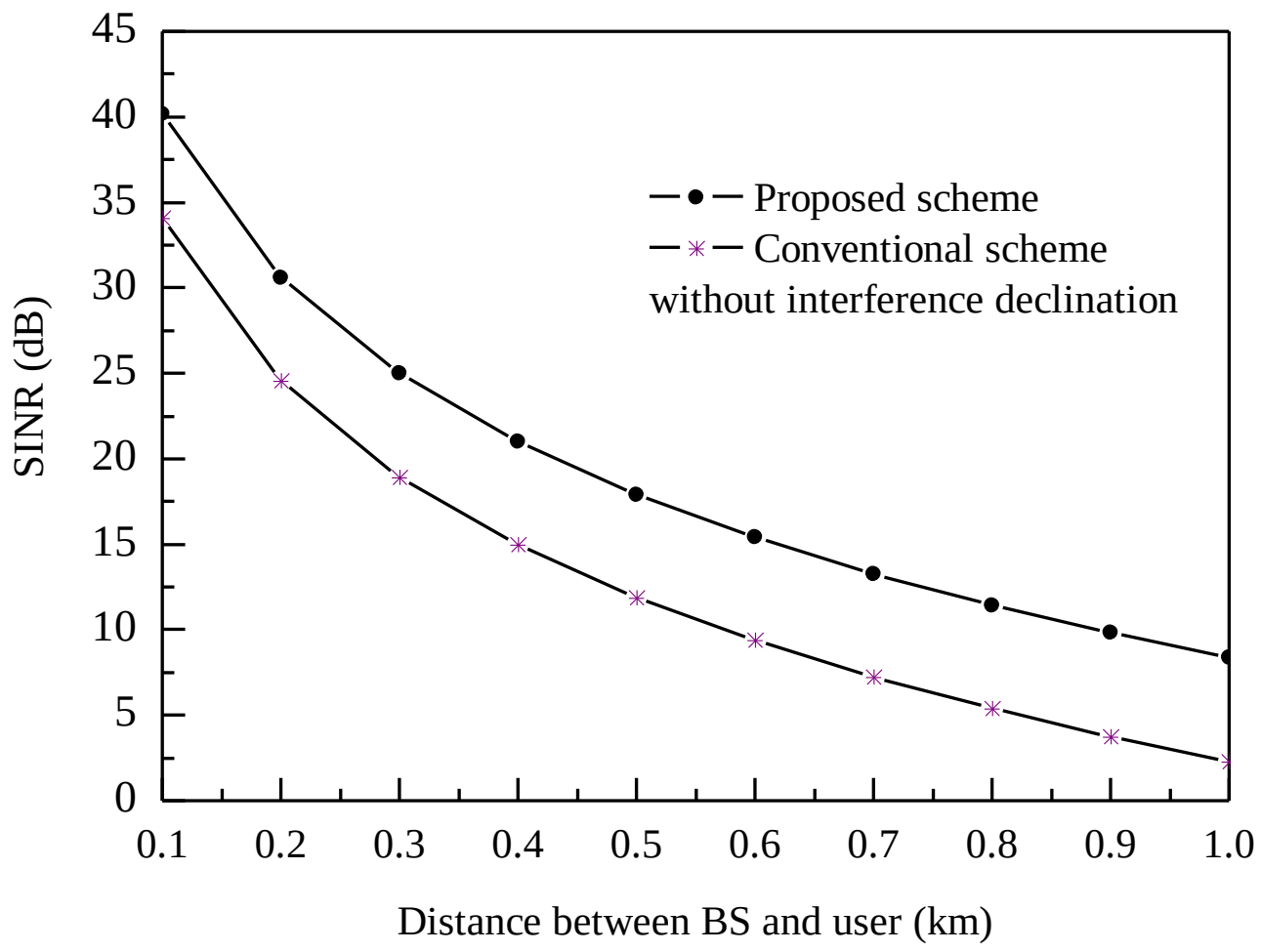

**Fig. 7.** Comparison of SINR levels in case of interference declination.

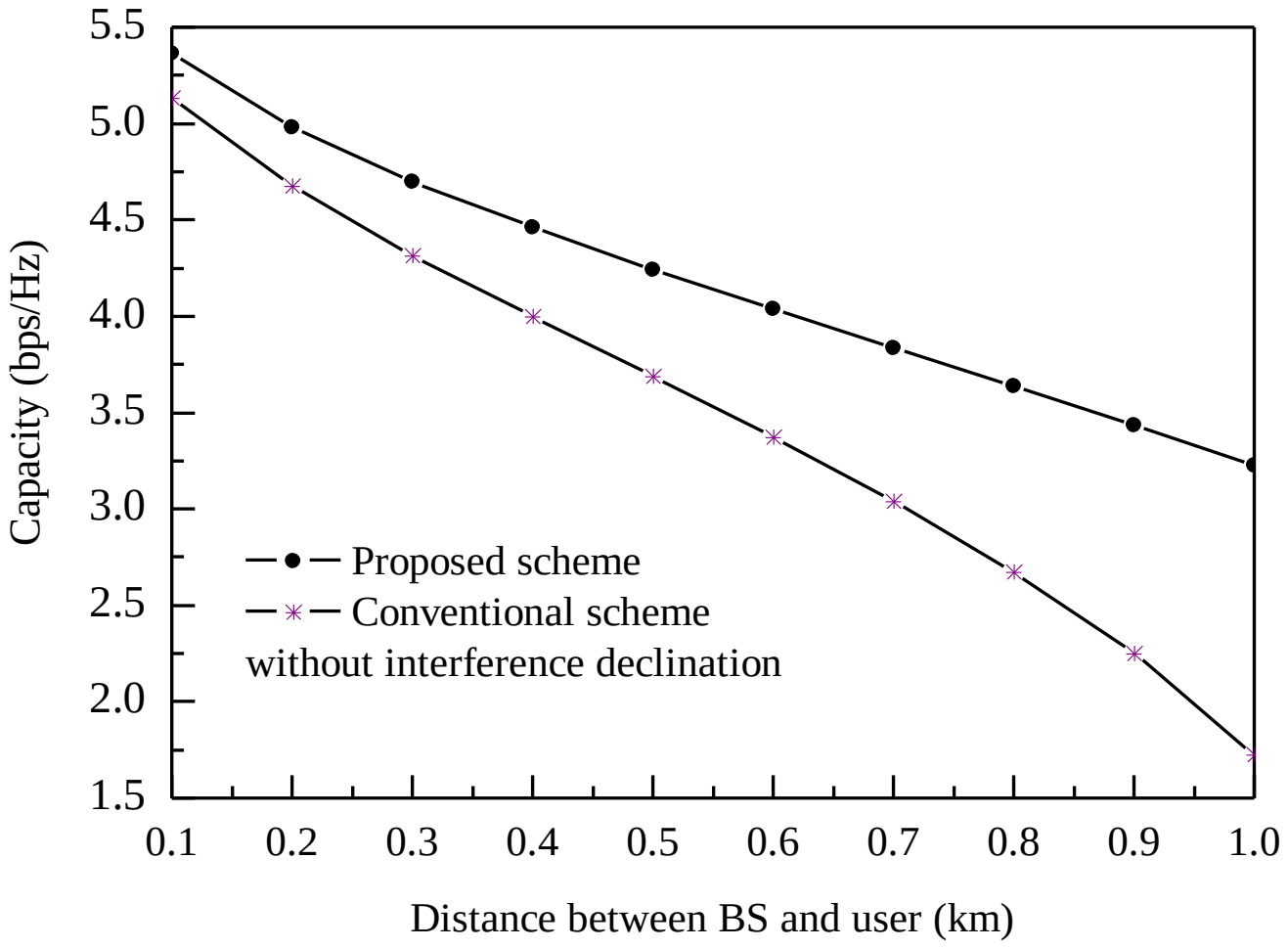

**Fig. 8.** Comparison of capacity in case of interference declination.

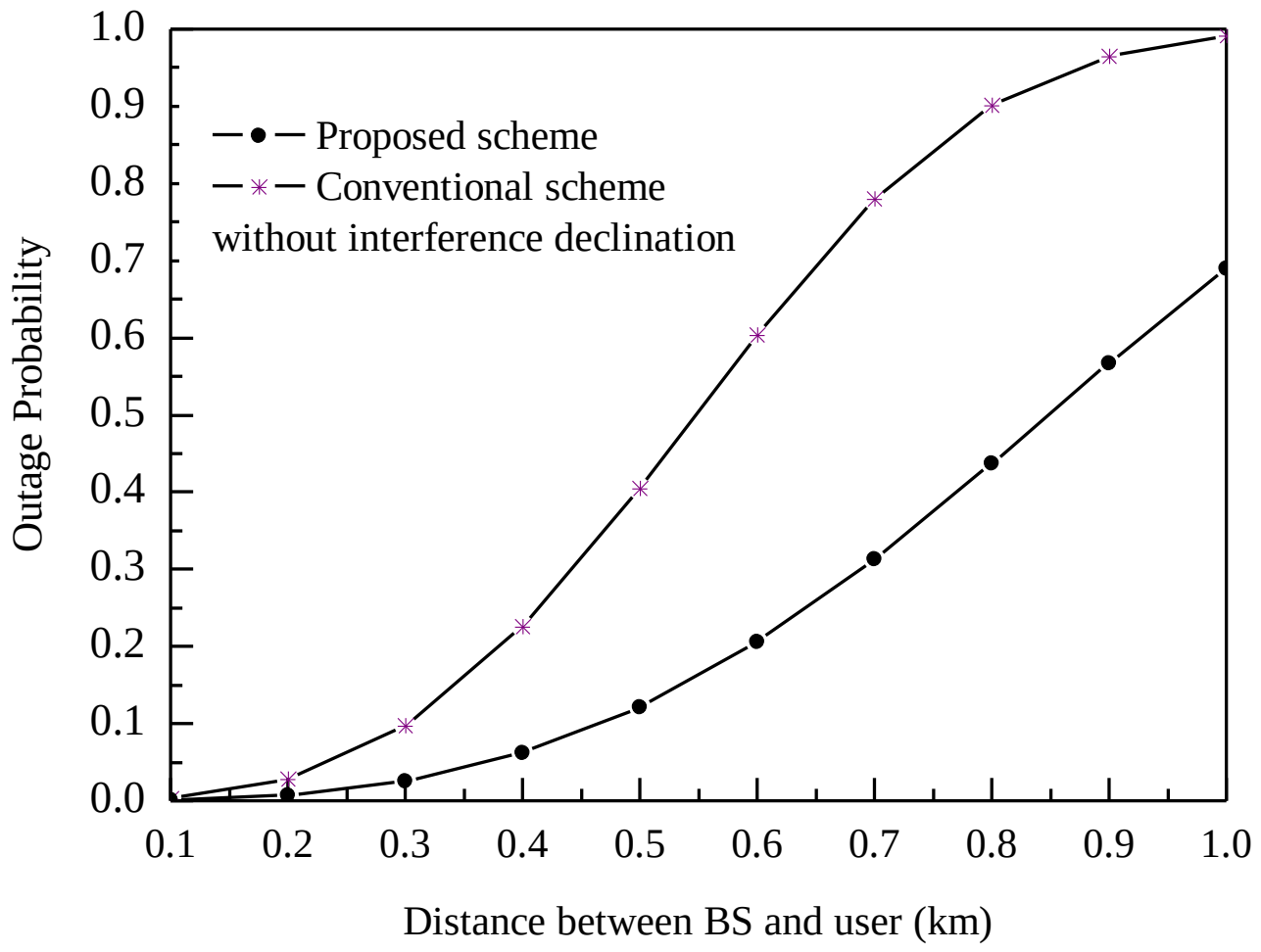

**Fig. 9.** Comparison of outage probability in case of interference declination.

## V. Conclusion

Our deep inspection perceives the interference declination for dynamic channel borrowing scheme. The main advantage of the proposed scheme is the usage of identical radio spectrum at the same time with inconsequential interference. Various situations have been illustrated for comfort of understanding so that it can fascinate the interest for the future wireless networks. In the earlier time, there were lots of researches concerning interference mitigation. Our proposed scheme shows two unique ways for interference management after the dynamic channel borrowing process. We demonstrate the performances of the proposed scheme which is satisfactory. Inter-cell interference declination, inter-carrier interference declination in OFDMA, interference declination in multi-cell networks are our next research content.

## References

[1] Mostafa Zaman Chowdhury, Yeong Min Jang, and Zygmunt J. Haas, "Call Admission Control Based on Adaptive Bandwidth Allocation for Wireless Networks," *IEEE/KICS Journal of Communications and Networks*, vol. 15, no. 1, pp. 15-24, February 2013.

[2] Ibrahim Habib, Mahmoud Sherif, Mahmoud Naghshineh, and Parviz Kermani, "An Adaptive Quality of Service Channel Borrowing Algorithm for Cellular Networks," *International Journal of Communication System,* vol. 16, no. 8, pp. 759–777, October 2003.

[3] Do Huu Tri, Vu Duy Loi, and Ha Manh Dao, "Improved Frequency Channel Borrowing and Locking Algorithm in Cellular Mobile Systems," In Proceeding of *IEEE International Conference on Advanced Communication Technology,* February 2009, pp. 214 – 217.

[4] Zuoying Xu, Pitu B. Mirchandani, and Susan H. Xu, "Virtually Fixed Channel Assignment in Cellular Mobile Networks with Recall and Handoffs," *Telecommunication Systems,* vol. 13, no. 2-4, pp. 413-439, July 2000.

[5] Geetali Vidyarthi, Alioune Ngom, and Ivan Stojmenovic´, "A Hybrid Channel Assignment Approach Using an Efficient Evolutionary Strategy in Wireless Mobile Networks," IEEE Transactions on Vehicular Technology, vol. 54, no. 5, pp. 1887-1895, September 2005.

[6] Kyuho Son and Song Chong, "Dynamic Association for Load Balancing and Interference Avoidance in Multi-Cell Networks," IEEE Transactions on Wireless Communications, vol. 8, no. 7, pp. 3566-3576, July 2009.

[7] Xunyong Zhang, Chen He, Lingge Jiang, and Jing Xu, "Inter-cell Interference Coordination Based on Softer Frequency Reuse in OFDMA Cellular Systems," In Proceeding of *IEEE International Conference Neural Networks & Signal Processing,* June 8, 2008, pp. 270-275.

[8] Xia Wang and Shihua Zhu, "Mitigation of Intercarrier Interference Based on General Precoder Design in OFDM Systems," In Proceeding of *International Conference on Advanced Information Networking and Applications*, May 2009, pp. 705-710.

[9] Mostafa Zaman Chowdhury, Yeong Min Jang, Choong Sub Ji, Sunwoong Choi, Hongseok Jeon, Junghoon Jee, and Changmin Park, "Interface Selection for Power Declination in UMTS/WLAN Overlaying Network," In Proceeding of *IEEE International Conference on Advanced Communication Technology*, February 2009, pp. 795-799.

[10] Mostafa Zaman Chowdhury, Yeong Min Jang, and Zygmunt J. Haas, "Cost-Effective Frequency Planning for Capacity Enhancement of Femtocellular Networks," *Wireless Personal Communications*, vol. 60, no. 1, pp. 83-104, September 2011.

[11] Shaoji Ni, Yong Liang, and Sven-Gustav Häggman, "Outage Probability in GSM-GPRS Cellular Systems with and without Frequency Hopping," *Wireless Personal Communication,* vol. 14, no. 3, pp. 215-234, September 2000.